# CPU Simulation with Ranked Set Sampling and Repeated Subsampling

Magnus Ekman
NVIDIA
mekman@nvidia.com

***Abstract*—Computer system simulation studies routinely rely on executing a limited number of short application regions, since full end-to-end simulation is prohibitively time-consuming. To preserve representativeness, existing methods employ either random sampling or phase-based characterization to identify representative regions.**

**In this work, we revisit random sampling in the context of computer architecture simulation. To assess how the confidence level varies with different micro-architectural configurations, we examine how the sample standard deviation relates to the sample mean. We show that the ranked set sampling (RSS) technique—well established in the statistical literature—maps naturally to architectural simulation and yields significantly tighter confidence intervals than simple random sampling. Across our experiments, RSS reduces the confidence interval width by up to 50%.**

**We further introduce a repeated subsampling scheme that identifies representative simulation regions for future studies. For a fixed sample size, this approach reduces the maximum observed error from 35% to 10%. Evaluating two selection criteria, we find that more informed subsample selection provides additional accuracy gains. Overall, our method achieves an average error below 2% and a maximum error of 3.5% across individual SPEC CPU 2017 Integer applications when simulating 30 regions of 1 million instructions each.**

## I. INTRODUCTION

End-to-end computer system simulation of modern applications is generally infeasible due to the slow execution speed of cycle-accurate performance simulators. As a result, most studies rely on simulating multiple short application regions rather than full program execution. Region selection was an active research area 20–25 years ago, leading to two dominant approaches: statistical sampling techniques [1][2][3][4][5] and phase-based characterization, best known through SimPoint [6][7][8]. Statistical sampling typically requires more simulation regions but provides confidence intervals to quantify accuracy. SimPoint usually selects far fewer regions but does not offer statistical confidence. Prior work has shown that per-application error can be large [2] and that increasing the number of SimPoints (up to ~300) can reduce this error and enable empirical confidence bounds [8].

Despite this, recent studies [9][10][11] and industry practice indicate that most evaluations use only 10–30 SimPoints per application, accepting potentially significant—and often unknown—error. A major reason is the high cost of warming (or maintaining state for) caches, prefetchers, and branch predictors. This practical constraint discourages sampling methods that require many simulation regions.

This work examines whether statistical sampling can be made practical with far fewer simulation regions while still delivering performance estimates with high and known accuracy. We first evaluate ranked set sampling (RSS) [12] as an alternative to simple random sampling and show that it works well for architectural simulation. We then introduce a repeated subsampling approach that selects a small, representative subset of simulation regions from a larger pool and quantifies the resulting error. Our study makes the following contributions:

1. **Application of RSS to architectural simulation.** We evaluate ranked set sampling in this context for the first time and show that it can reduce confidence interval width by up to 50% compared to simple random sampling at the same sample size.
2. **Repeated subsampling for region selection.** Starting from a larger set of simulated regions, we repeatedly draw smaller subsamples to identify a representative subset. Across all SPEC CPU 2017 Integer applications, this reduces the maximum observed error from 35% to 10% compared to simple random sampling with an equal number of regions.
3. **Refined selection criterion for improved accuracy.** We evaluate enhanced subsample selection and achieve an average error below 2%—and a maximum of 3.5%—when simulating only 30 regions.
4. **Characterization of sampled distributions.** We analyze performance variation across simulator configurations and show a linear relationship between CPI variance and mean. We further examine how simulator configuration changes affect performance of different regions, providing intuition for why RSS is effective in this domain.

The rest of this paper is organized as follows. Section II discusses sources of randomness in architectural simulation. Section III introduces ranked set sampling. Section IV describes our experimental methodology, and Section V presents results. Section VI discusses implications, and Section VII relates our findings to prior work. Section VIII concludes.

## II. Randomness in Computer System Simulations

Statistical sampling involves probability distributions, random variables, and associated means and standard deviations. A common goal is to estimate a confidence interval through repeated experiments. Applying these ideas to computer system simulation can appear contradictory, especially when using deterministic simulators that always produce identical results. To clarify this apparent dichotomy, we outline four distinct but related problems involving sampling and confidence interval estimation in architecture simulation:

1. **Estimating whole-application performance by simulating only selected regions.**
   Studies often simulate a subset of short regions and compute overall performance as a (potentially weighted) mean of their measured performance. Regions may be selected using random or systematic sampling [1][2][3] or through profiling-based phase identification such as SimPoint [6][7][8]. In sampling terminology, each region is a sampling unit; we refer to them as *simulation regions*.
2. **Ensuring valid thread interleavings when sampling multiprogram or multithreaded workloads.**
   Fast-forwarding to regions of interest is usually performed using a functional ISA-level model. In multicore systems, however, fixed-rate functional fast-forwarding may produce unrealistic interleavings. For example, when simulating a dual-core system, fast-forwarding both cores to instruction **N** implicitly assumes equal instruction throughput, even if one core would progress twice as fast in reality. This issue has been studied in [13][14], where detailed simulation and identified phases are used to guide fast-forwarding to produce representative interleavings.
3. **Measuring performance in multicore systems when instructions per cycle (IPC) is not a reliable proxy.**
   While IPC is meaningful for single-thread workloads, total instruction count may increase when an application is parallelized across cores—for instance, due to spinning on barriers. This is fine when measuring the full application, but sampled IPC can no longer directly quantify work completed. Instead, the simulator must track application-level progress, such as completed transactions [15][16]. This introduces a granularity challenge: if the sampling unit is short relative to the unit of work, quantization error increases. Thus, sampling units must be sufficiently long to capture meaningful progress.
4. **Evaluating system performance in the presence of execution noise.**
   Real systems are not performance-deterministic across repeated runs because starting conditions differ—interrupt timing, thread scheduling, cache content, and other microarchitectural states vary between invocations. Execution time can therefore be viewed as a random variable with a fixed mean and non-zero standard deviation. Even if an architectural enhancement reduces the mean execution time, individual runs may occasionally perform worse due to variance. Prior work quantified this effect through repeated simulation [15][16], using deliberate perturbations to emulate real-system variability. The authors labeled this *space variability* to distinguish it from *time variability* described in problem (1) above.

This work focuses exclusively on improving problem (1) in the context of single-core simulations.

### A. Using Sampling to Estimate Whole-Application Performance

To estimate full-application execution time using only sampled regions, prior work [1][2][3] applies classical statistical sampling theory to compute both an estimated mean cycles per instruction (CPI) [1] and a confidence interval (CI):

$$CI = \bar{y} \pm ME, \tag{1}$$

where $ME$ is the margin of error. Sample size is typically chosen to provide high confidence (e.g., 95%) that the margin of error remains small (e.g., 3%). Since simulators are deterministic, the CI reflects randomness in how regions are selected—not randomness in simulation outcomes. Interpreting the CI correctly is therefore critical: if many independent sampling experiments were performed, each selecting regions randomly, then in 95% of the experiments the estimated mean would lie within 3% of the true mean.

The reasoning above appears to imply that simulated performance for a given application will be incorrect 5% of the time. In practice, simulation regions are selected only once and are later reused, meaning the CI describes the probability of selecting a *representative set of regions*. Thus, for roughly 5% of the applications the selected regions will not be sufficiently representative, leading to consistently inaccurate performance estimates for those applications in all subsequent studies. Formally, the confidence interval is:

$$CI = \bar{y} \pm z_{\alpha/2}\frac{s}{\sqrt{n}}, \tag{2}$$

where $z_{\alpha/2}$ depends on the chosen confidence level, $s$ is the population standard deviation, and $n$ is the sample size. For a fixed confidence level (e.g., 95%), three factors determine whether the margin of error remains below 3% of the mean: sample size $n$, standard deviation $s$, and mean $\bar{y}$. In typical architectural studies, the same regions are used to evaluate a baseline and an alternative configuration, making $n$ constant. If the new feature affects performance, $\bar{y}$ will change. Whether $s$ changes depends on how uniformly the feature affects different regions. If the micro-architectural feature affects only some of the regions, then $s$ will change [2], while if it affects all regions by the same amount, $s$ will stay constant. However, if $\bar{y}$ increases as well, a larger $s$ does not necessarily require a larger $n$ to maintain less than 3% error, because the CI is relative to the mean.

[1] We mainly report CPI, since its use allows arithmetic mean across regions. With a fixed instruction count per region, IPC would require a harmonic mean—analogous to averaging speeds over equal distances.

[2] It decreases when affected regions move toward the mean and increases when they move farther from it.

To examine this behavior empirically, we simulated all ten SPEC CPU 2017 Integer applications across multiple simulator configurations. For each application, we collected enough regions to produce tight confidence intervals (Section IV describes configurations and sample sizes). We computed the mean CPI and corresponding standard deviation. Fig. 1 plots standard deviation versus mean CPI. Each point represents one simulator configuration.

The data shows an approximately linear relationship between standard deviation and mean, though slopes differ by application and may be flat or slightly negative. This suggests that standard deviation does not fluctuate arbitrarily across configurations—supporting the interpretation that randomness primarily arises from region selection, not configuration changes.

Fig. 2 reports the resulting 95% confidence intervals when applying simple random sampling with $n = 100$ regions, expressed as margin of error. For instance, for perlbench Config 0, repeating the sampling experiment many times would result in a sampled mean within 14% of the true mean in 95% of trials.

The width of each confidence interval is determined by the ratio of standard deviation to mean—consistent with the scatter patterns in Fig. 1. For mcf, this ratio is nearly constant across configurations, yielding similar margins of error. For xalancbmk, the ratio varies significantly, producing nearly a 3× difference in margin of error between Config 0 and Config 6.

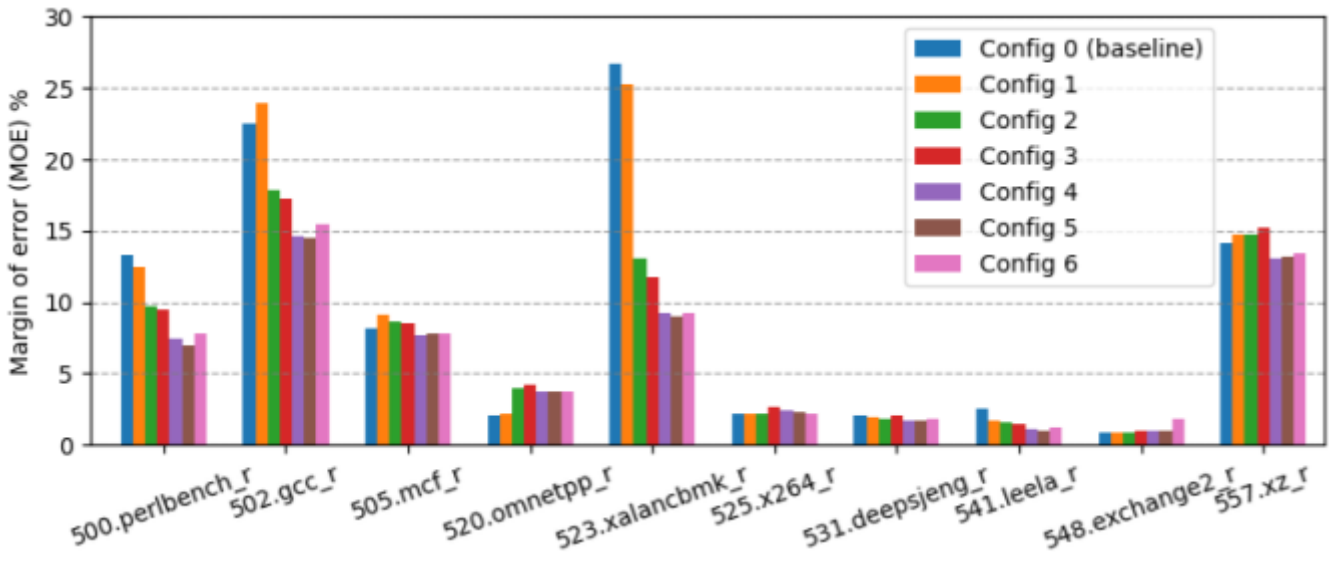

Fig. 2. Margin of error resulting from simple random sampling with n=100 at 95% confidence.

## III. Ranked Set Sampling

In light of these observations, we now describe a statistical sampling technique known as ranked set sampling (RSS) and why it is well suited to computer system simulations. This technique was first introduced in 1952 [12], with renewed interest in recent years [17][18]. The procedure is illustrated in Fig. 3 and proceeds as follows:

1. Define two parameters, $M$ and $K$, where $M$ is the number of cycles[3] and $K$ is the number of sets (and the set size). Randomly select $M \times K$ sets, each containing $K$ sampling units, yielding $MK^2$ units in total.
2. Within each set, order the $K$ units based on an approximation of their values.
3. For the first cycle, select the smallest unit from set 0, the second smallest from set 1, and so on, until one unit has been chosen from each of the first $K$ sets. Repeat this process for the remaining sets to complete $M$ cycles.
4. Use the resulting $MK$ units as the final sample and measure the target metric.

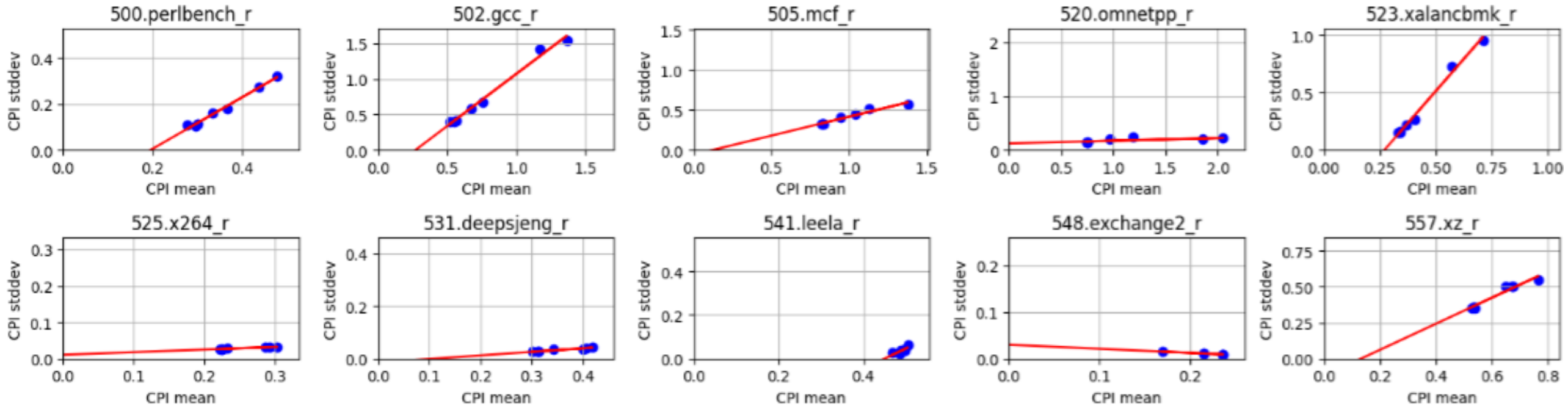

Fig. 1. Standard deviation as a function of mean CPI across SPEC 2017 Integer applications.

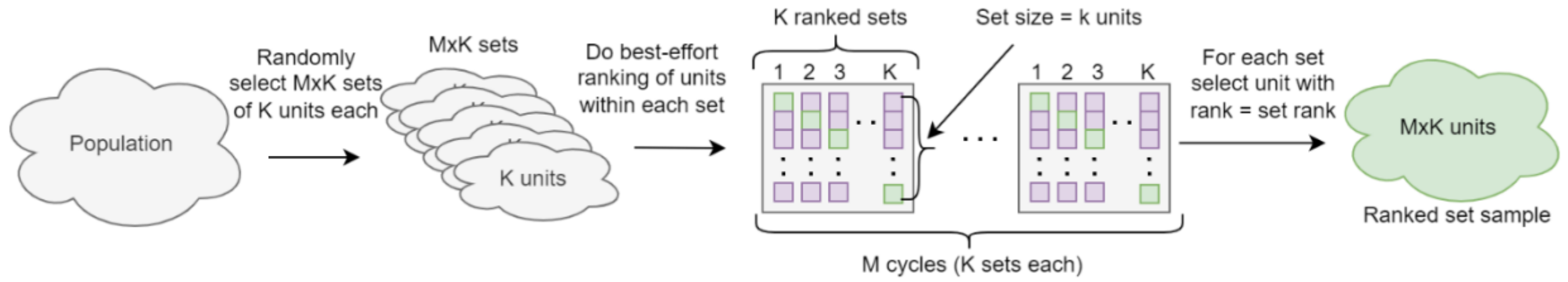

Fig. 3. Illustration of the ranked set sampling process.

[3] This does not refer to simulated cycles.

To illustrate with an example distinct from computer simulation, consider estimating the average height of trees in a forest. With $M = 1$, we would randomly select $K$ sets of $K$ trees each. Trees in each set are ranked visually—an inexpensive approximation for height. We then select one ranked tree from each set following step 3, measure only those $K$ trees precisely, and compute the mean height.

With accurate ranking, RSS yields a tighter confidence interval than simple random sampling with the same final sample size. Importantly, the estimator remains unbiased even with imperfect ordering [19], though the benefit diminishes as ordering accuracy degrades. In the limiting case of random ranking, RSS converges to simple random sampling.

### *A. Ranked Set Sampling for Computer System Simulation*

To apply RSS in architectural simulation, we begin by randomly selecting $MK^2$ simulation regions. Each region is simulated once using a baseline configuration, and the resulting CPI values are used to rank the regions within each set. Following the RSS selection process described above, we choose $MK$ regions—these become the sample used in subsequent simulation studies. Fig. 4 illustrates the process for $M = 2$ and $K = 3$.

Two aspects of this workflow may initially appear counterintuitive. First, ordering regions requires detailed simulation rather than a quick performance estimate. Second, simulating $MK^2$ regions but using only $MK$ regions to compute the mean seems inefficient. The first issue is explained by the fact that ordering regions based on a baseline configuration serves as an approximation of ordering for other configurations: while some regions may shift slightly in relative performance, ordering is generally expected to remain similar. In terms of number of simulated regions, although RSS requires simulating $MK^2$ regions up front this cost is amortized across all future studies that use only the selected $MK$ regions. The approach deliberately invests initial effort to select a representative and well-distributed set of regions. Even if the baseline configuration later changes, the selected $MK$ regions can continue to be reused, and RSS still offers tighter confidence intervals than selecting $MK$ regions randomly—assuming ordering remains approximately valid within each set. This assumption is studied empirically in Section V.

## IV. Methodology

To quantify the accuracy of RSS, we conducted experiments using a cycle-accurate simulator modeling a single ARM v9 CPU core. We defined a baseline microarchitectural configuration (Config 0) and six progressively higher-performance alternatives (Config 1–Config 6). The baseline is an out-of-order, four-wide core with modest cache capacities, a basic stream prefetcher, and a TAGE branch predictor [20]. Table I summarizes all configurations, where highlighted cells indicate modified parameters for each configuration. These changes span multiple parts of the core, including the branch predictor, reorder buffer capacity, pipeline width, cache sizes, latencies, and the addition of a spatial memory streaming (SMS) prefetcher [21] and a best-offset (BO) prefetcher [22].

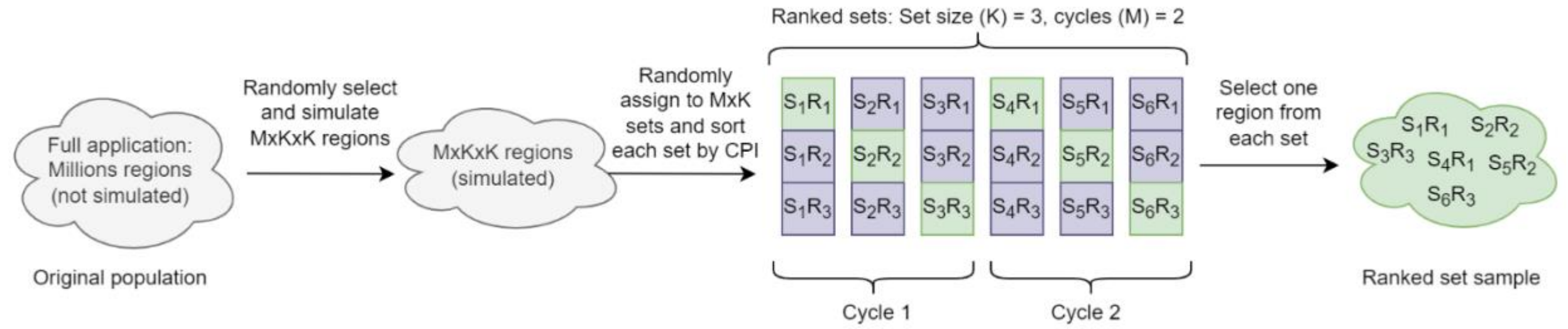

Fig. 4. Process for applying ranked set sampling to computer system simulation.

TABLE I. Parameters for the simulated configurations.

| | Config 0 | Config 1 | Config 2 | Config 3 | Config 4 | Config 5 | Config 6 |
|---|---|---|---|---|---|---|---|
| Fetch width | 8 | 8 | 8 | 8 | 8 | 8 | 8 |
| Issue width | 8 | 8 | 8 | 8 | 8 | 8 | 8 |
| D-cache hit latency | 3 | 3 | 3 | 3 | 3 | 3 | 3 |
| L2 hit latency | 8 | 8 | 8 | 8 | 8 | 8 | 8 |
| I-cache size | 32KB | 64KB | 64KB | 64KB | 64KB | 64KB | 64KB |
| D-cache size | 32KB | 64KB | 64KB | 64KB | 64KB | 64KB | 64KB |
| L2 cache size | 512KB | 1M | 1M | 1M | 1M | 1M | 1M |
| L3 cache size | 2MB | 4MB | 4MB | 4MB | 4MB | 4MB | 4MB |
| SMS PF | Disabled | Disabled | Enabled | Enabled | Enabled | Enabled | Enabled |
| Reorder buffer size | 128 | 128 | 128 | 256 | 256 | 256 | 256 |
| Physical regs | 128 | 128 | 128 | 256 | 256 | 256 | 256 |
| Retire width | 4 | 4 | 4 | 8 | 8 | 8 | 8 |
| Memory latency | 130ns | 130ns | 130ns | 130ns | 90ns | 90ns | 90ns |
| L3 hit latency | 30ns | 30ns | 30ns | 30ns | 20ns | 20ns | 20ns |
| L2 best-offset PF | Disabled | Disabled | Disabled | Disabled | Disabled | Enabled | Enabled |
| TAGE BPU tables | 4 | 4 | 4 | 4 | 4 | 4 | 8 |
| TAGE table entries | 2048 | 2048 | 2048 | 2048 | 2048 | 2048 | 4096 |

We evaluated the SPEC CPU 2017 Integer Rate benchmark suite, compiled with gcc 12. For experimental reference, we first generated a large pool of simulation regions for each configuration. Each region consisted of one million instructions, following sufficient warm-up to populate caches and other microarchitectural state.

For each application, we selected close[4] to 1,000 simulation regions unless a larger number was required to ensure a tight confidence interval under simple random sampling. This allowed us to treat the sample mean from the full region pool as the ground-truth performance estimate. The resulting region counts are listed in Table II.

TABLE II. NUMBER OF SIMULATED REGIONS PER APPLICATION.

| Application | Simulation regions (sample size) |
|---|---|
| 500.perlbench_r | 1,997 |
| 502.gcc_r | 6,195 |
| 505.mcf_r | 964 |
| 520.omnetpp_r | 967 |
| 523.xalancbmk_r | 6,861 |
| 525.x264_r | 915 |
| 531.deepsjeng_r | 1,041 |
| 541.leela_r | 1,062 |
| 548.exchange2_r | 1,030 |
| 557.xz_r | 3,047 |

For the sampling experiments in Section V we used a sample size of 30, which is commonly considered sufficient for reliable confidence interval estimation [23]. Sampling was done by selecting data points (simulation regions) from the large number of data points in Table II.

## V. Experimental Results

Fig. 5 reports the IPC for each application and simulator configuration. The confidence intervals—shown as small error bars—are extremely narrow, confirming that the values can serve as ground-truth references when evaluating sampling schemes with smaller sample sizes. The overall geometric mean performance ranges from 1.52 to 2.56, meaning Config 6 delivers 68% higher performance than the baseline (Config 0). We intentionally selected configurations spanning this wide range to evaluate both incremental and more disruptive architectural changes.

### *A. Ranked Set Sampling*

To compare RSS with simple random sampling (SRS) used in prior work [1][2][3], we repeated each sampling experiment 1,000 times, each using a different seed to select simulation regions. Fig. 6 shows the resulting distributions of sampled means for a sample size of 30. Because both sampling approaches are unbiased, the distributions are centered near the true mean.

Although rare, some sampled means deviate substantially from the true mean. RSS reduces the likelihood of such extreme outcomes, resulting in a more concentrated distribution. Note that ranking was performed using Config 0, while accuracy was evaluated using Config 6 for both RSS and SRS—reflecting the effect of ranking not perfectly transferring across configurations.

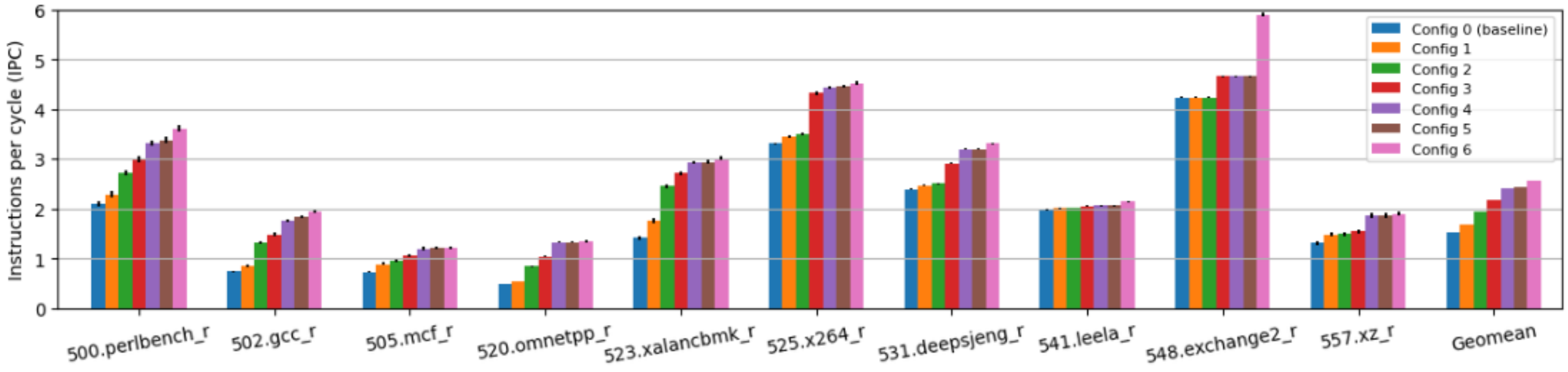

Fig. 5. IPC and confidence intervals for each application and configuration, using the full set of simulation regions listed in Table II.

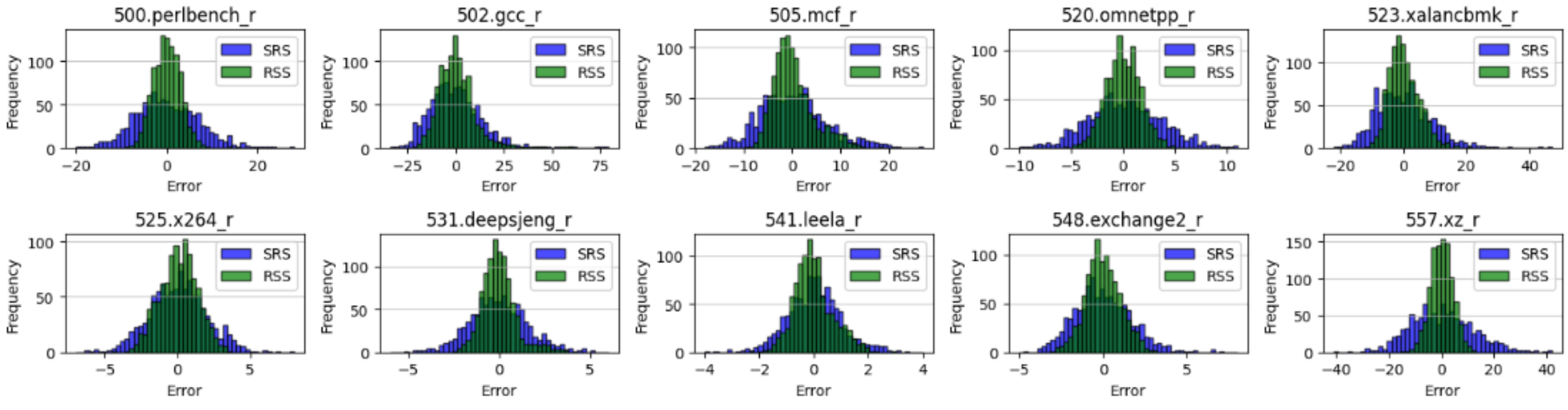

Fig. 6. Distribution of sampled means from 1,000 experiments (sample size = 30). RSS produces a noticeably tighter distribution.

[4] The exact number of collected regions varied due to infrastructure-specific implementation details.

To quantify the benefit of RSS over SRS, we derived empirical 95% confidence intervals based on the range containing 95% of samples. We evaluated RSS for multiple values of $M$(1, 2, and 3), keeping the total sample size fixed at 30 regions. Fig. 7 compares analytical and empirical confidence intervals for SRS and the empirical intervals for RSS.

The analytical SRS intervals closely match the empirical intervals, although they are slightly conservative. All RSS configurations yield tighter confidence intervals than SRS, with $M = 1$ performing best—indicating that ranking accuracy is high[5].

To examine ranking quality more directly, we analyzed the selected regions for $M = 1$ across all configurations and visualized the true ranking of each selected region within its set (Fig. 8). Each chart contains one curve for each simulator configuration. The x-axis represents the $K$ sets and the y-axis the true ranking for the selected sampling unit for each set. Perfect ranking appears as the line $y = x$, which holds for Config 0 by construction. While other configurations deviate slightly, the ordering remains similar, explaining why RSS performs well across configurations.

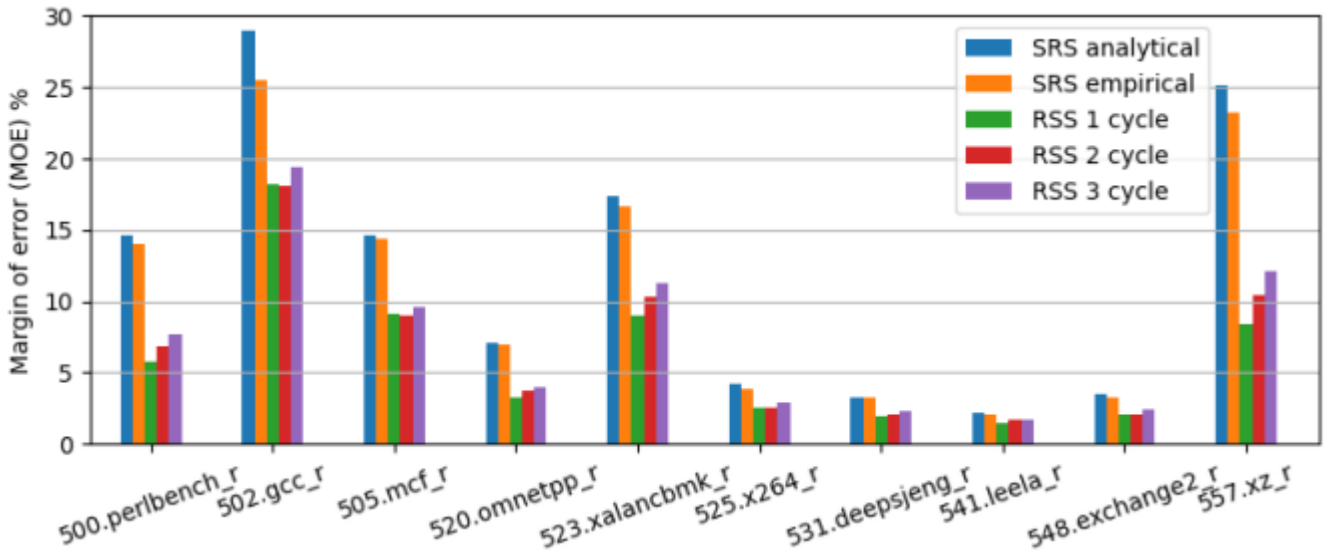

Fig. 7. Empirical and analytical confidence intervals for different sampling schemes (sample size = 30).

## B. *Improved Accuracy with Repeated Subsampling*

Although RSS reduces variance, Fig. 6 and Fig. 7 show that a single sample of 30 regions may still occasionally produce a mean far from the true value. However, the distributions in Fig 6 also suggest a remedy: perform repeated sampling and select the subsample whose mean is closest to the true mean. In fact, because we already simulated a large number of regions (Table II) we have a reliable estimate of the true mean. We can therefore draw many subsamples—each of size 30—and select the one whose mean is closest to the full-sample mean. This dramatically reduces the error compared to relying on an arbitrary sample resulting from a single trial.

This approach can be applied to both SRS and RSS. Two challenges arise. First, we do not have a closed-form analytical confidence interval for the resulting estimator. Second, the subsample selected based on Config 0 may not generalize to other simulator configurations. However, based on Section II.A and results in Section V.A, a subsample close to the baseline mean is unlikely to fall at a distribution tail for another configuration. We evaluate this empirically using the flow shown in Fig. 9.

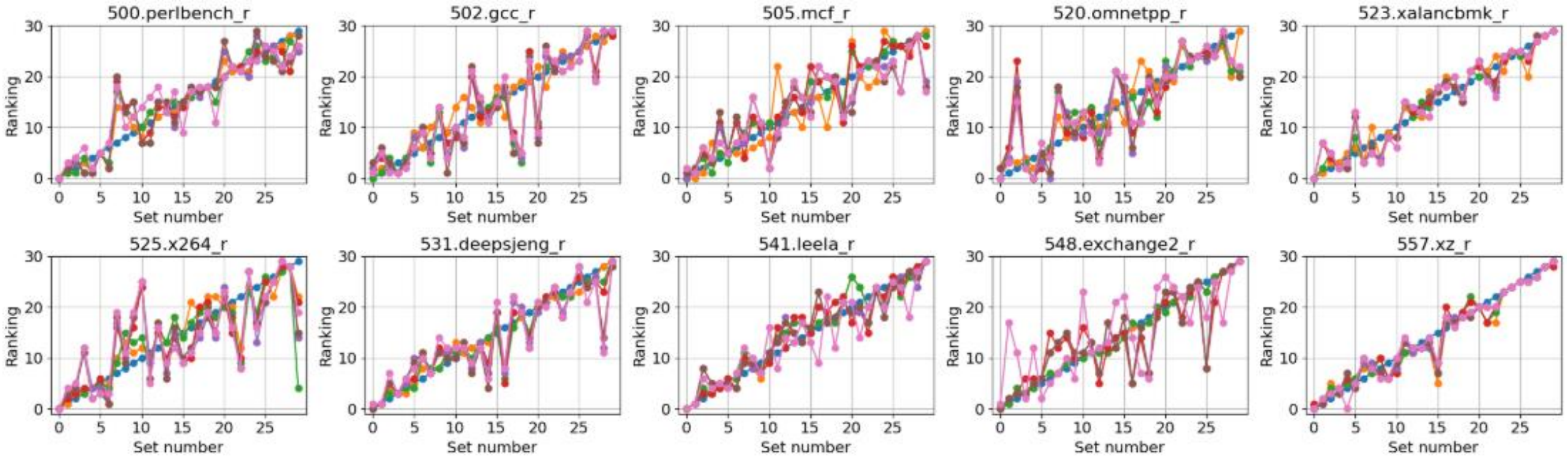

Fig. 8. Visualization of ranking accuracy across configurations.

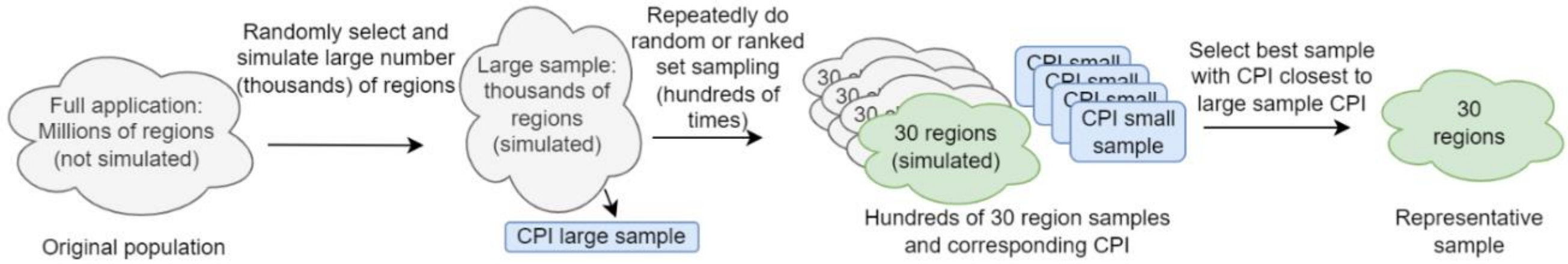

Fig. 9. Experimental flow for evaluating repeated subsampling.

[5] For ranked set sampling, a larger value of M is used when the ranking accuracy is low.

1. Randomly select and simulate a large pool of regions to obtain an accurate CPI estimate.
2. Repeatedly draw subsamples of size 30 using either SRS or RSS.
3. Compute each subsample's mean CPI and compare it to the accurate estimate from (1).
4. Select the subsample from step (2) with the mean closest to the accurate estimate.

We performed 1,000 repeated sampling trials for both SRS and RSS, each with a sample size of 30. For each technique, we selected the subsample whose mean was closest to the true mean (based on Config 0), then evaluated its accuracy across the remaining configurations (Config 1 – Config 6). For comparison, we also evaluated SRS and RSS when sampling only once, yielding four different schemes: SRS once, RSS once, SRS repeated, and RSS repeated. The results are shown in Fig. 10. Each data point represents the measured error (not confidence interval) for a given configuration.

As expected, sampling only once can lead to large errors. For example, SRS exceeds 20% error for two gcc configurations and reaches 35% for xz. RSS performs better, but still exceeds 20% error for xalancbmk in one configuration. These observations align with the wide confidence intervals in Fig. 7.

Repeated subsampling dramatically improves accuracy. Errors are below 10% in all cases—and often in the low single digits. Interestingly, SRS performs comparably to RSS, suggesting that the RSS benefit largely disappears when repeated subsampling is applied.

### C. *Modified Selection Criterion for Better Generalization*

We next investigated whether subsample selection could be improved by considering multiple configurations rather than only the baseline. Although this approach requires simulating all regions for multiple configurations, it may lead to better generalization. We modified the process as follows: Instead of requiring the subsample mean to be close to the true mean only for the baseline configuration, we selected a subsample considering three configurations (Config 0–Config 2). That is, instead of comparing two scalars (the mean for Config 0) between the subsample and the large sample, we now compared two vectors (the means for Config 0–Config 2). We selected the subsample that minimized the Chebyshev distance—the maximum element-wise difference—between the subsample mean vector and the true mean vector across the three configurations[6]. Fig. 11 illustrates the concept.

At the far left of the figure are the accurate means computed from the full set of simulation regions. We treated the first three configurations as training data used to select the best subsample, and the remaining configurations as test data used to evaluate generalization. The figure shows three candidate subsamples and their corresponding errors relative to the accurate means. Within the training set, subsample 3 is the best choice because it yields the smallest maximum error (4%, compared to 8% and 7%). We would therefore select subsample 3 and then examine its errors for the test configurations (Config 3–Config 6). In this example, the largest observed test error is 5% for Config 3. Fig. 12 presents the results using real simulation data. Only four data points appear for each sampling technique because the first three configurations (Config 0–Config 2) are used for subsample selection.

The two leftmost bars in each group correspond to baseline-only selection and match the repeated subsampling results in Fig. 10. For applications such as gcc, mcf, and xalancbmk, using multiple configurations during selection significantly reduces error (note that the y-range is different between Fig. 10 and Fig. 12). In most cases, errors fall below 2%, and all remain below 3.5%. As before, RSS provides no additional benefit over SRS when repeated subsampling is applied.

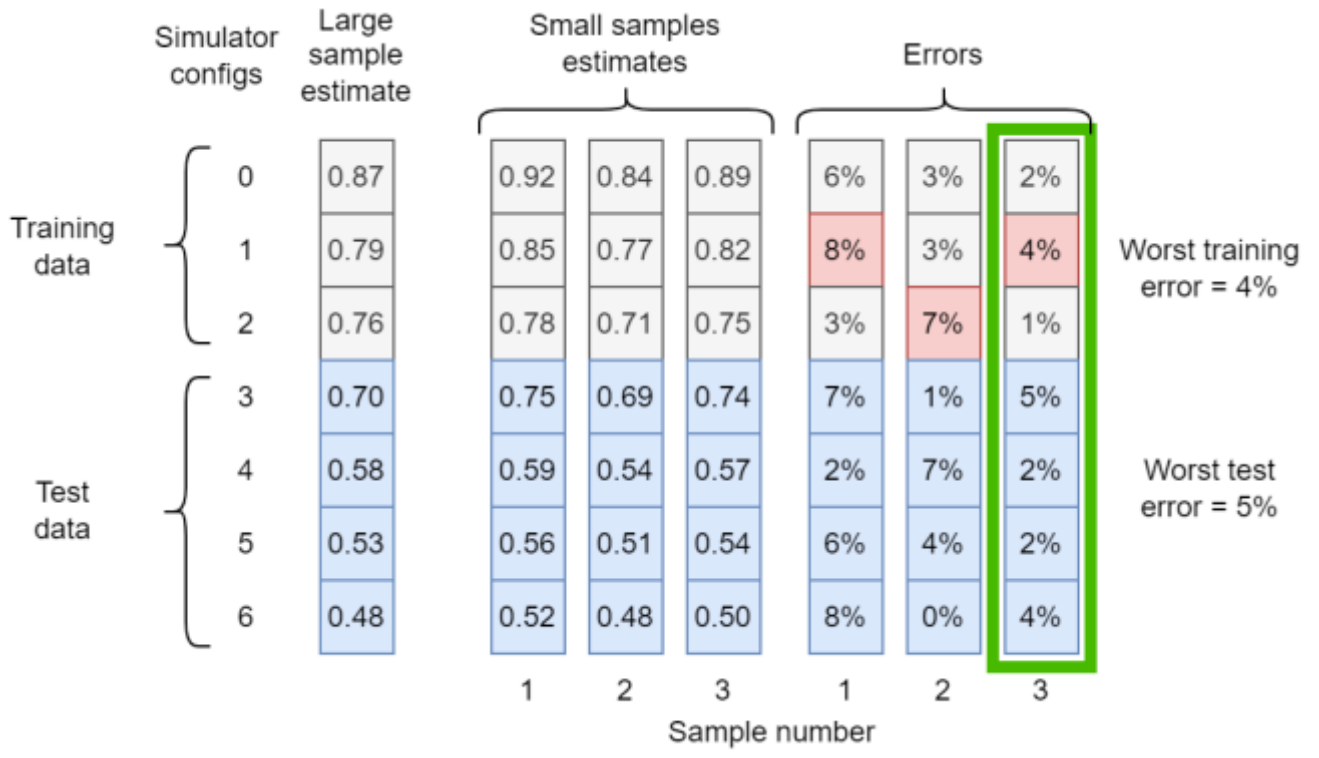

Fig. 11. Figure 1. Selecting subsamples by minimizing Chebyshev distance across multiple configurations.

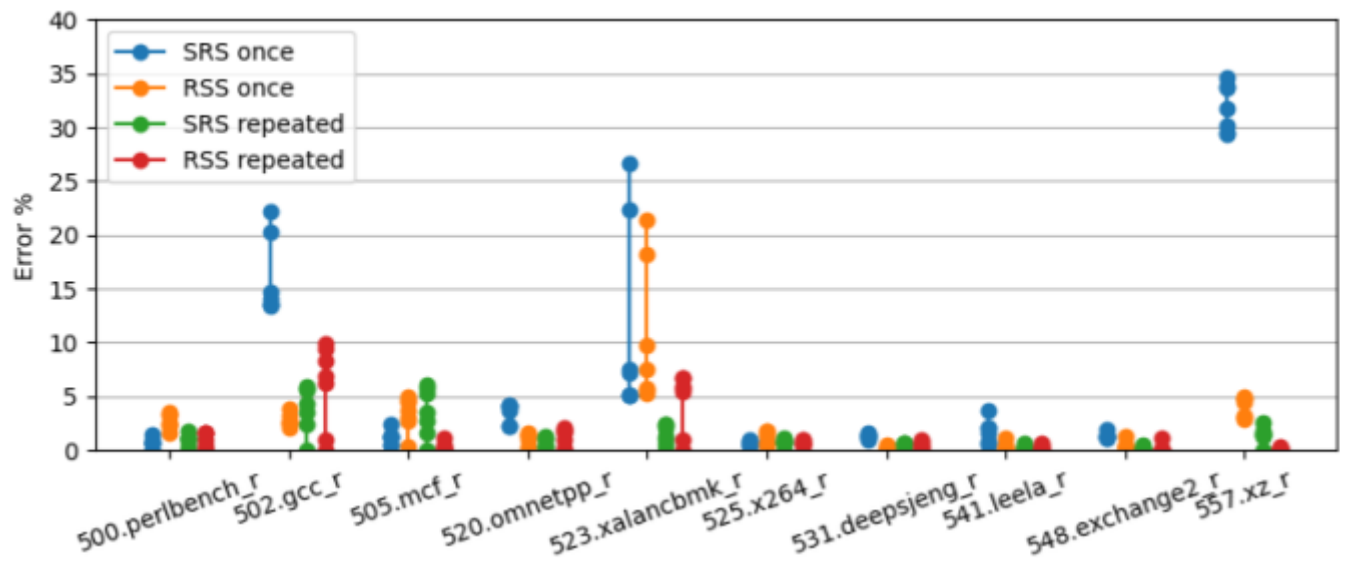

Fig. 10. Measured errors for the different sampling schemes (observed errors, not confidence intervals).

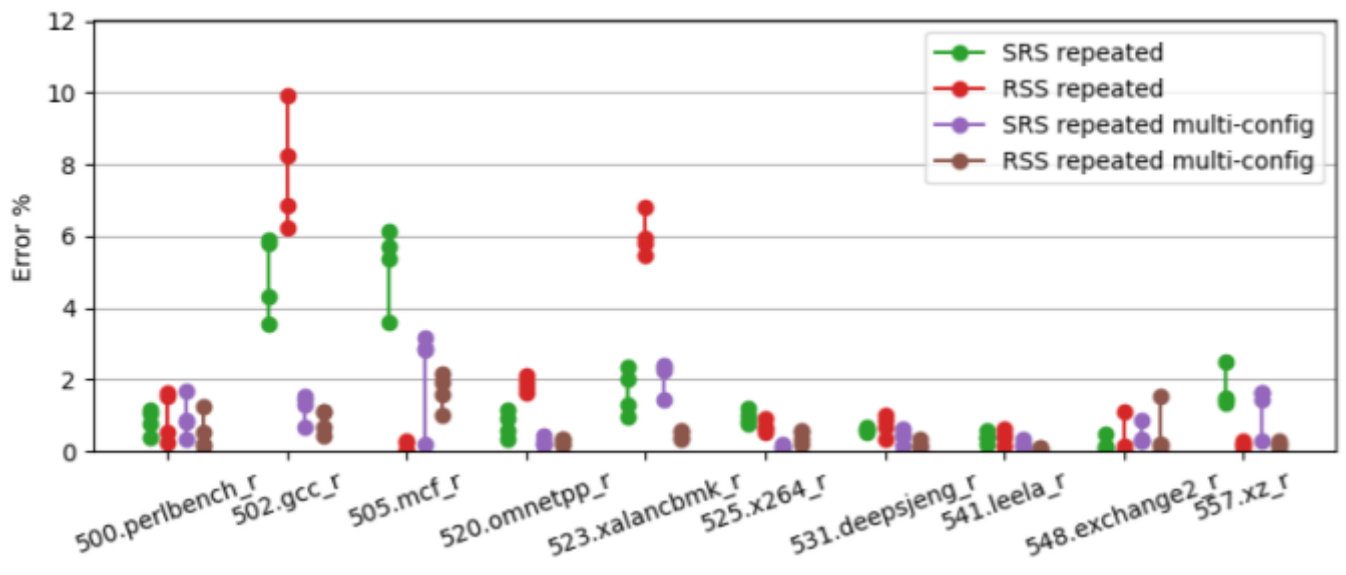

Fig. 12. Measured errors using different selection criteria.

---

[6] We also explored alternative criteria, e.g., maximizing correlation between the two vectors.

## VI. Discussion

Section V compared the sampling efficiency of ranked set sampling (RSS) against simple random sampling (SRS), followed by an evaluation of repeated subsampling to select a representative subset of simulation regions from a larger pool. While the empirical results demonstrate low error across simulator configurations, another consideration is the resulting statistical validity of these approaches—particularly with respect to confidence intervals and potential sampling bias.

### A. *Confidence Interval and Sample Bias*

In order to determine if our proposed techniques result in in an unbiased estimate and if we can quantify the confidence interval we need to discuss the three cases presented in Sections V.A, V.B, and V.C separately since the three schemes differ with respect to confidence intervals and bias.

For Section V.A, where SRS and RSS are compared directly, prior statistical results show that both methods yield unbiased estimators of the mean and support confidence interval computation [19][23]. For RSS, interval tightness depends on ranking accuracy, which we quantified empirically in Fig. 7 and showed that it was reduced by more than 50% in some cases. An additional insight from Fig. 1 is that standard deviation correlates strongly with the mean, enabling prediction of approximate required sample sizes without explicitly computing variance for the sample.

Repeated subsampling (Section V.B) raises additional considerations. We are unaware of a closed-form analytical confidence interval for the resulting estimator. Regarding bias, each individual subsample is an unbiased estimator of the mean; any bias must therefore arise from the selection criterion. For example, if the selection criterion was to select the subsample with the lowest mean, the resulting mean would be biased to be lower than the true mean. With our selection criterion—minimizing the distance to the true mean—it stands to reason that the subsample would be biased toward the true mean, i.e., it is unbiased. However, other statistics such as sample variance could still be biased. Since our objective was accurate mean CPI estimation, we did not investigate secondary effects further.

When the selection criterion is extended using Chebyshev distance across multiple configurations (Section V.C), unbiasedness with respect to the baseline mean cannot be guaranteed. This raises an important point. Our objective is not to provide an accurate estimate for the baseline configuration, since our methodology assumes that we already have an accurate estimate for that based on the large number of initial simulations. Our goal is to estimate the performance of *other* configurations by using far fewer regions. The implications of this are discussed next.

### B. *Accuracy for Other Configurations*

Fig. 12 shows that selecting small subsamples based on accuracy across three configurations (Config 0–Config 2) yields low error on the remaining configurations (Config 3–Config 6) in practice, even though the small sample size cannot provide narrow statistical error bounds. This problem is therefore no longer purely a statistical sampling problem but involves evaluating whether a characterization of simulation regions generalizes across architectures. This resembles a machine learning (ML) problem where we want to establish how well a model based on one data set generalizes to a different data set [24]. That is also how we treated the problem in Section V.C, where we used Config 0–Config 2 to *train* the model (identify what simulation regions to use) and then used Config 3–Config 6 to *evaluate* or *test* the model (quantify the accuracy resulting from selected regions).

As with ML generalization, guarantees are limited. Good performance on the evaluated configurations does not ensure accuracy for architectural designs that differ substantially from those seen during selection. Nonetheless, it is reasonable to expect high accuracy for configurations exhibiting similar behavior, which is typically the case unless the microarchitecture is completely different than the baseline configuration.

### C. *Other Considerations*

The primary cost of the proposed approach is the initial detailed simulation of a large region pool. Repeated subsampling itself is essentially free, since all candidates are drawn from existing simulation results. Although the upfront cost may appear large, repeated subsampling results in a strict reduction in simulation cost compared to SRS, which requires the large region pool to be simulated for every experiment. In contrast, repeated subsampling only incurs this upfront burden once and it is amortized across future architectural studies. Nevertheless, the cost remains nontrivial. Two strategies may mitigate it. First, selecting subsamples using only one configuration, as in Section V.B, may be sufficient in many cases—Fig. 10 shows only a few modest outliers. Second, the initial large-region simulations might be performed using a faster, less detailed simulator. While this would not yield an accurate reference mean, it may still enable identification of representative subsamples. We view this as a promising direction for future work.

A notable drawback of repeated subsampling relative to SRS or RSS is the absence of a quantified confidence interval for the final estimate. Although empirical results suggest low error, some users may require formal bounds. One practical mitigation is to periodically evaluate additional simulation regions to verify that the chosen subsample remains representative under evolving microarchitectural assumptions.

Finally, although our experiments focused on single-core simulations, the technique itself is not inherently single-core. As discussed in Section II, there are additional, but orthogonal, challenges associated with sampling performance in multi-core systems. Exploring if existing solutions for those problems can be combined with the repeated subsampling technique remains future work.

## VII. Related Work

The work most closely related to this study includes [1][2][3], which apply SRS to estimate full-application performance. Two additional studies [4][5] use SRS to estimate the performance difference between two simulator configurations rather than absolute performance. In contrast,

Section V.A of this paper evaluates RSS, which provides higher accuracy for the same sample size.

SRS and RSS are only two examples of statistical sampling techniques. Another example is stratified sampling [23], which previously has been studied in the context of computer architecture simulation [26][27][28]. It differs from SRS and RSS by requiring the population to first be stratified into disjoint strata based on an ancillary variable.

This paper introduces repeated subsampling to identify a small set of simulation regions that yields more accurate estimates than a single random sample of equal size. Both repeated subsampling and RSS rely on the observation that simulation regions producing accurate estimates for one configuration often remain accurate for others, consistent with observations in [4][5][25] that estimating performance differences tends to incur less error than estimating absolute performance.

Both RSS and repeated subsampling invest upfront effort to characterize simulation regions before selecting those to simulate further. This idea is related to SimPoint [6][7][8], though implemented differently. SimPoint relies on microarchitecture-independent basic block vectors (BBVs) to identify representative program phases, whereas RSS and repeated subsampling use simulated IPC from a baseline configuration. As a result, the techniques in this study can account for variation in data access behavior not captured by BBVs.

As described in Section II, sampling has been used to address issues related to multicore simulation such as finding valid thread-interleavings [13][14] and measuring performance in presence of timing-induced noise [15][16]. These are distinct goals from ours.

Ranked set sampling was introduced in 1952 [12] and later shown to yield unbiased estimators of the mean [19]. It has since been applied in several scientific domains [17][18], but to our knowledge has not previously been applied to computer system simulation.

The repeated subsampling technique proposed here does not seem to correspond to an established sampling method in other fields. It resembles resampling-based approaches [30][31], with bootstrapping being the most familiar example. However, such techniques are typically used to estimate properties of the underlying distribution by using all subsamples, whereas our use case selects a single subsample from repeated draws.

## VIII. Conclusion

A central challenge in computer system simulation is estimating full-application performance while simulating only a small subset of the application. Sampling-based approaches address this challenge, but traditional techniques often require more simulation regions than are practical. This work evaluated ranked set sampling (RSS) in this context and, to our knowledge, represents its first application to architectural simulation. RSS reduced confidence interval width by up to 50% compared to simple random sampling with the same sample size.

We then introduced repeated subsampling, in which a large initial pool of simulated regions is used to select a smaller, representative subset. This approach reduced the maximum observed error from 35% to 10% while requiring only 30 simulation regions per SPEC CPU 2017 Integer application. Extending the selection criterion to consider multiple configurations further improved accuracy, yielding an average error below 2% and a maximum error of 3.5%.

Beyond evaluating these techniques, we characterized sampled distributions across several simulator configurations and showed that CPI variance scales linearly with the mean. This relationship enables prediction of required sample sizes for desired confidence intervals. We also demonstrated that region ordering from one configuration often transfers to others, supporting reuse of selected subsets across studies.

Historically, the large sample sizes needed to achieve tight confidence intervals have limited the practical adoption of statistically grounded sampling, leading to uncertainty in experimental results. This study addresses this problem by providing two insights. First, applying repeated subsampling allows for selecting a small sample that is expected to closely match a larger sample. Second, the validity of this smaller sample can be examined periodically, or after making major microarchitectural changes by rerunning the larger sample. The latter of these insights is quite obvious but surprisingly powerful.